\documentclass[aps,twocolumn,tightenlines]{revtex4}
\usepackage[dvipdf]{graphicx}

\begin{document}
\title{Measurement of the electric polarizability of lithium by atom interferometry}

\author{A. Miffre, M. Jacquey, M. B\"uchner, G. Tr\'enec and J. Vigu\'e}
\address{ Laboratoire Collisions Agr\'egats R\'eactivit\'e -IRSAMC
\\Universit\'e Paul Sabatier and CNRS UMR 5589
 118, Route de Narbonne 31062 Toulouse Cedex, France
\\ e-mail:~{\tt jacques.vigue@irsamc.ups-tlse.fr}}

\date{\today}

\begin{abstract}

We have built an atom interferometer and, by applying an electric
field on one of the two interfering beams, we have measured the
static electric polarizability of lithium  $\alpha =(24.33 \pm
0.16)\times10^{-30} $ m$^3$ with a $0.66$\% uncertainty. Our
experiment is similar to an experiment done on sodium in 1995 by
D. Pritchard and co-workers, with several improvements: the
electric field can be calculated analytically and our phase
measurements are very accurate. This experiment illustrates the
extreme sensitivity of atom interferometry: when the atom enters
the electric field, its velocity increases and the fractional
change, equal to $4 \times10^{-9}$ for our largest field, is
measured with a $10^{-3}$ accuracy.

\end{abstract}
\maketitle

\begin{figure}
\includegraphics[width = 6 cm,height= 4.5 cm]{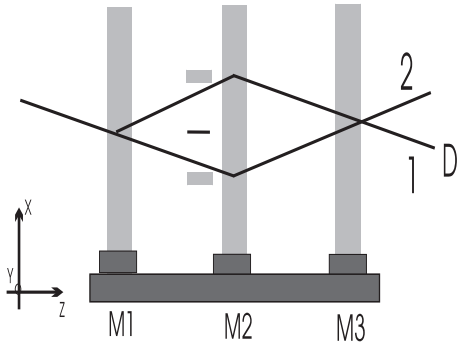}
\caption{\label{interferometer} Schematic drawing of our
Mach-Zehnder atom interferometer: a collimated atomic beam, coming
from the left, is diffracted by three laser standing waves and the
output beam $1$ selected by a slit is detected by a hot-wire
detector D. The capacitor with a septum is placed just before the
second laser standing wave. The $x$,$y$ and $z$ axis are defined.}
\end{figure}


An atom interferometer is the ideal tool to measure any weak
modification of the atom propagation due to electromagnetic or
inertial fields. The application of a static electric field is
particularly interesting because it gives access to the electric
polarizability $\alpha$ and this quantity cannot be measured by
spectroscopy which is sensitive only to polarizability differences
(for a review on polarizability measurements, see reference
\cite{bonin97}).

Several experiments with atom interferometers have exhibited a
sensitivity to the electric electric field
\cite{shimizu92,nowak98,nowak99} without aiming at a
polarizability measurement and interferometers using an inelastic
diffraction process have been used to measure the difference of
polarizability between the ground state and an excited state
\cite{rieger93,morinaga96}. A very accurate measurement of the
atom polarizability $\alpha$ requires that a well-defined electric
field is applied on only one interfering beam and, up-to-now, such
an experiment has been made only by D. Pritchard et al.
\cite{ekstrom95,roberts04} by inserting a thin electrode, a
septum, between the two atomic paths. We have made a similar
experiment with our lithium atom interferometer, represented in
figure \ref{interferometer} and we are going to describe its first
results. With respect to the experiment of D. Pritchard et al., we
have made several improvements: we have designed a capacitor with
an analytically calculable electric field; we have a better phase
sensitivity; finally our interferometer based on laser diffraction
is species selective. Our experimental accuracy is presently
limited by the knowledge of the mean atom velocity.

When an electric field $E$ is applied, the ground state energy
decreases by the polarizability term $U = - 2 \pi \epsilon_0
\alpha E^2$. Therefore, when an atom enters the electric field,
its kinetic energy increases and its wave vector $k$ becomes $ k
+\Delta k$, with $\Delta k = 2 \pi \epsilon_0 \alpha E^2 m/(\hbar
k)$. The resulting phase shift $\phi $ of the atomic wave is given
by:

\begin{equation}
\label{n2} \phi = \frac{2 \pi \epsilon_0 \alpha }{\hbar v} \int
E^2(z) dz
\end{equation}

\noindent $v = \hbar k/m$ is the atom velocity and the spatial
dependence of the electric field along the atomic path is taken
into account.

\begin{figure}
\includegraphics[width = 7 cm,height= 3 cm]{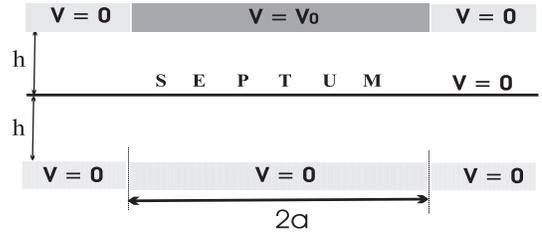}
\caption{\label{capacitor} Schematic drawing of the capacitor. The
septum is parallel to the $z$-axis and the electrodes are located
at $x= \pm h \approx 2$ mm. The high voltage electrodes at the
potential $V_0$ extends from $z=-a$ to $z=+a$, while the guard
electrodes extend outside with $|z|>a$, with $a \approx 25$ mm.
The septum and the guard electrodes are at $V=0$. }
\end{figure}

To know precisely the electric field along the atomic path, guard
electrodes are needed, as discussed in \cite{ekstrom95}. We have
developed a capacitor where guard electrodes are in the plane of
the high voltage electrode, as shown in figure \ref{capacitor},
which defines the notations. In this case, the field can be
expressed analytically from the potential distribution $V(z,x=h)$
in the plane of the high-voltage electrode. We give here only the
results of the calculation which will be published elsewhere
\cite{miffre05a}.

The integral of $E^2$ along the septum surface can be written :

\begin{equation}
\label{n3} \int E(z,0)^2 dz = \left[\frac{V_0}{h}\right]^2 L_{eff}
\end{equation}

\noindent $V_0/h$ is the electric field of an infinitely long
capacitor and the capacitor effective length $L_{eff}$ is given
by:

\begin{equation}
\label{n31} L_{eff} \approx 2a - (2h/\pi)
\end{equation}

\noindent where exponentially small corrections of the order of
$\exp(-2\pi a/h)$ are neglected. The atoms do not sample the
electric field on the septum surface but at a small distance $x$
from the septum and we should add to the effective length a small
correction proportional to $x^2$. In our experiment, with $x
\lesssim 50$ $\mu$m and $h\approx 2$ mm, this correction is below
$10^{-4} L_{eff}$ and negligible.

The capacitor external electrodes are made of thick glass plates
covered by an aluminium layer. The guard electrodes are insulated
from the high voltage electrode by $100$ $\mu$m wide gaps which
have been made by laser evaporation and, under vacuum, we can
operate the capacitor up to $V = 450$ V. The glass spacers are
glued on the external electrodes and the septum, made of a $6$
$\mu$m thick mylar foil aluminized on both faces, is stretched and
glued on the electrode-spacer assemblies. In our calculation, we
assume that the potential on the high-voltage electrode is known
everywhere but we ignore the potential inside the $100$ $\mu$m
wide dielectric gaps which may get charged. A superiority of our
design is that these gaps are very narrow, thus minimizing the
uncertainty on the capacitor effective length. Another defect is
that the spacer thicknesses are not perfectly constant. We use
equation (\ref{n3}) by replacing $h$ by its mean value
$\left<h\right>$, thus making a relative error of the order of
$\left<(h -\left <h\right> )^2\right> / \left<h\right>^2$ which is
fully negligible.

We have previously described our three-grating Mach-Zehnder atom
interferometer \cite{delhuille02a,miffre05}. The lithium atomic
beam is a supersonic beam seeded in argon and we use Bragg
diffraction on laser standing waves at $\lambda = 671$ nm. By
choosing a laser detuned by about $3$ GHz on the blue side of the
$^2S_{1/2}$ - $^2P_{3/2}$ transition of the $^7Li$ isotope, the
signal is almost purely due to this isotope (natural abundance
$92.4$\%) and not to the other isotope $^6Li$. Any other species
present in the beam, for instance lithium dimers or heavier alkali
atoms, is not diffracted and does not contribute to the signal. In
three-grating interferometers, the phase of the interference
fringes depends on the $x$-position of the gratings depending
themselves on the position $x_i$ of the mirrors $M_i$ forming the
three laser standing waves and this phase is given by $\psi = 2p
k_L (x_1+x_3-2x_2)$, where $k_L$ is the laser wavevector and $p$
is the diffraction order. By scanning the position $x_3$ of mirror
$M_3$, we have observed interference fringes with an excellent
visibility ${\mathcal{V}}$, up to $84.5$\%.

\begin{figure}
\includegraphics[width = 7 cm,height= 5.5 cm]{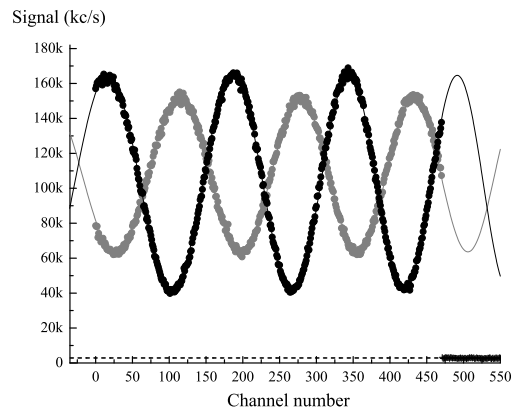}
\caption{\label{signals} Experimental signals and their fits (full
curves) corresponding to $V=0$ (black dots) and $V_0 \approx 260$
Volts (grey dots): the phase shift is close to $3\pi$ with a
reduced visibility.}
\end{figure}

The capacitor is placed just before the second laser standing
wave, with the septum between the two atomic beams. In the present
work, we have used only the diffraction order $p=1$ so that the
center of the two beams are separated by about $90$ $\mu$m in the
capacitor. When the septum is inserted between the two atomic
paths, the atom propagation is almost not affected and we observe
interference fringes with a visibility $\mathcal{V} = 84$ \% and a
negligible reduction of the atomic flux. To optimize the phase
sensitivity, we have opened the collimation slit $S_1$ and the
detection slit $S_D$ (see reference \cite{miffre05}) with widths
$e_1= 18$ $\mu$m and $e_D = 50$ $\mu$m, thus increasing the mean
flux up to $10^5$ counts/s and slightly reducing the fringe
visibility down to ${\mathcal{V}_0} = 62$\% (see figure
\ref{signals}). We have made a series of recordings, labelled by
an index $i$ from $1$ to $44$, with $V_0=0$ when $i$ is odd and
with $V_0 \neq 0$ when $i$ is even  with $V_0 \approx 10\times i $
Volts. For each recording, we apply a linear ramp on the
piezo-drive of mirror $M_3$ in order to observe interference
fringes and $471$ data points are recorded with a counting time
per channel equal to $0.36$ s. Figure \ref{signals} presents a
pair of consecutive recordings. The high voltage power supply has
stability close to $ 10^{-4}$ and the applied voltage is measured
by a HP model 34401A voltmeter with a relative accuracy better
than $10^{-5}$.

The data points $I_i(n)$ have been fitted by a function $I_i(n) =
I_{0i} \left[1 + {\mathcal{V}}_i\cos \psi_i (n) \right]$, with
$\psi (n)= a_i + b_i n + c_i n^2$ where $n$ labels the channel
number, $a_i$ represents the initial phase of the pattern, $b_i$
an ideal linear ramp and $c_i$ the non-linearity of the
piezo-drive. For the $V=0$ recordings, $a_i$, $b_i$ and $c_i$ have
been adjusted as well as the mean intensity $I_{0i}$ and the
visibility ${\mathcal{V}}_i$, while, for the $V \neq 0$ recording,
we have fitted only $a_i$, $I_{0i}$ and ${\mathcal{V}}_i$, while
fixing $b_i$ and $c_i$ to their value $b_{i-1}$ and $c_{i-1}$ from
the previous $V=0$ recording. Our best phase measurements are
given by the mean phase $\left< \psi_i \right>$ obtained by
averaging $\psi_i (n)$ over the $471$ channels. The $1\sigma$
error bar of these mean phases are of the order of $2-3$ mrad,
increasing with the applied voltage up to $23$ mrad because of the
reduced visibility.

The mean phase values $\left< \psi_i \right>$ values of the
$V_0=0$ recordings present a drift equal to $ 7.5 \pm 0.2 $
mrad/minute and some scatter around this regular drift. The drift
is explained by the differential thermal expansion of the
structure supporting the three mirrors: its temperature was
steadily drifting at $1.17 \times 10^{-3}$ K/minute during the
experiment. We have no explanation of the phase scatter, which
presents a quasi-periodic structure: its rms value is equal to
$33$ milliradian and unfortunately this error dominates our phase
determination.

\begin{figure}
\includegraphics[width = 7 cm,height= 6.5 cm]{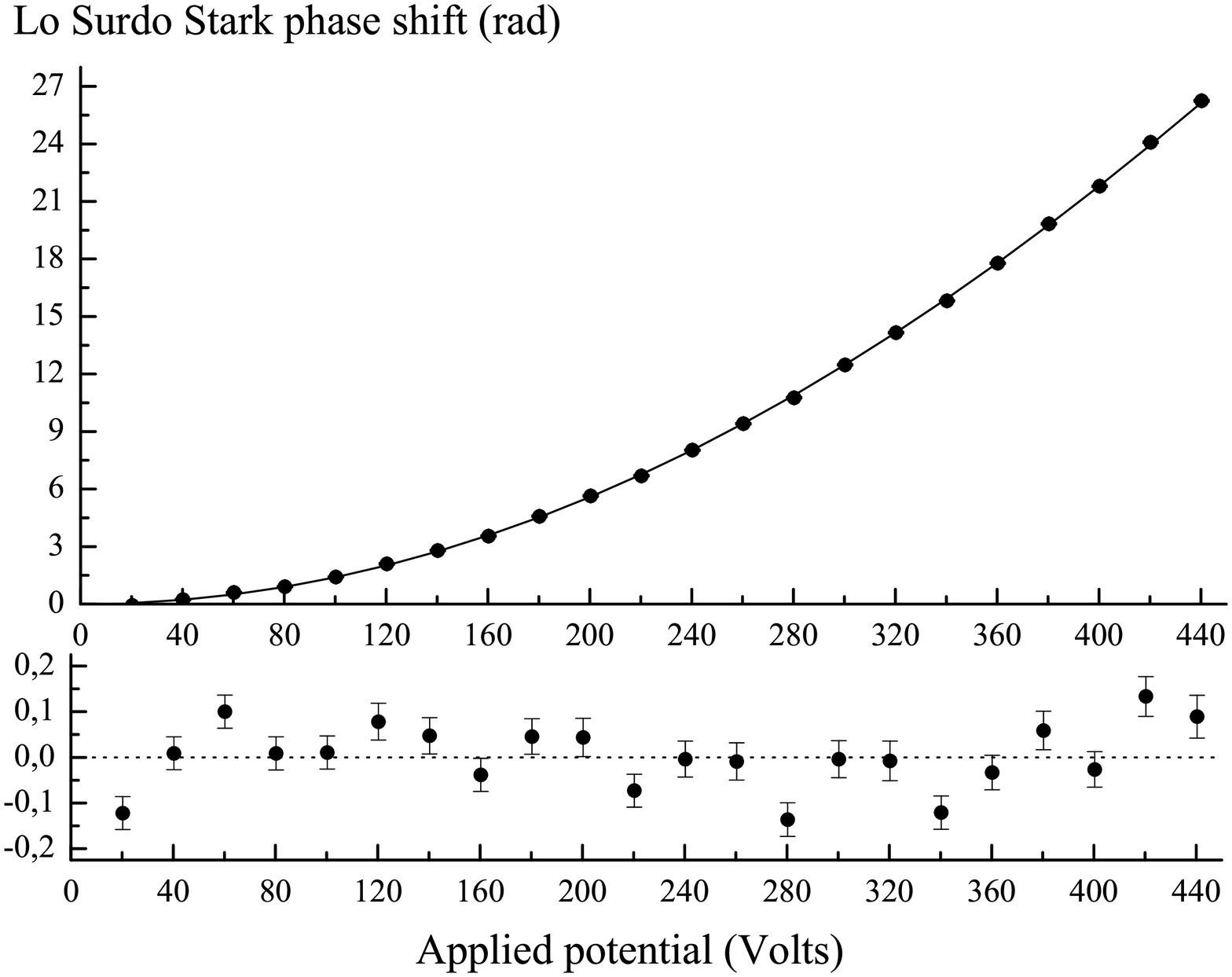}
\caption{\label{phaseshifts} Phase shift $\phi(V_0)$ as a function
of the applied voltage $V_0$: the best fit using equations
(\ref{n4},\ref{n5}) is represented by the full curve and the
residuals are plotted in the lower graph.}
\end{figure}

\begin{figure}
\includegraphics[width = 7 cm,height= 4.5 cm]{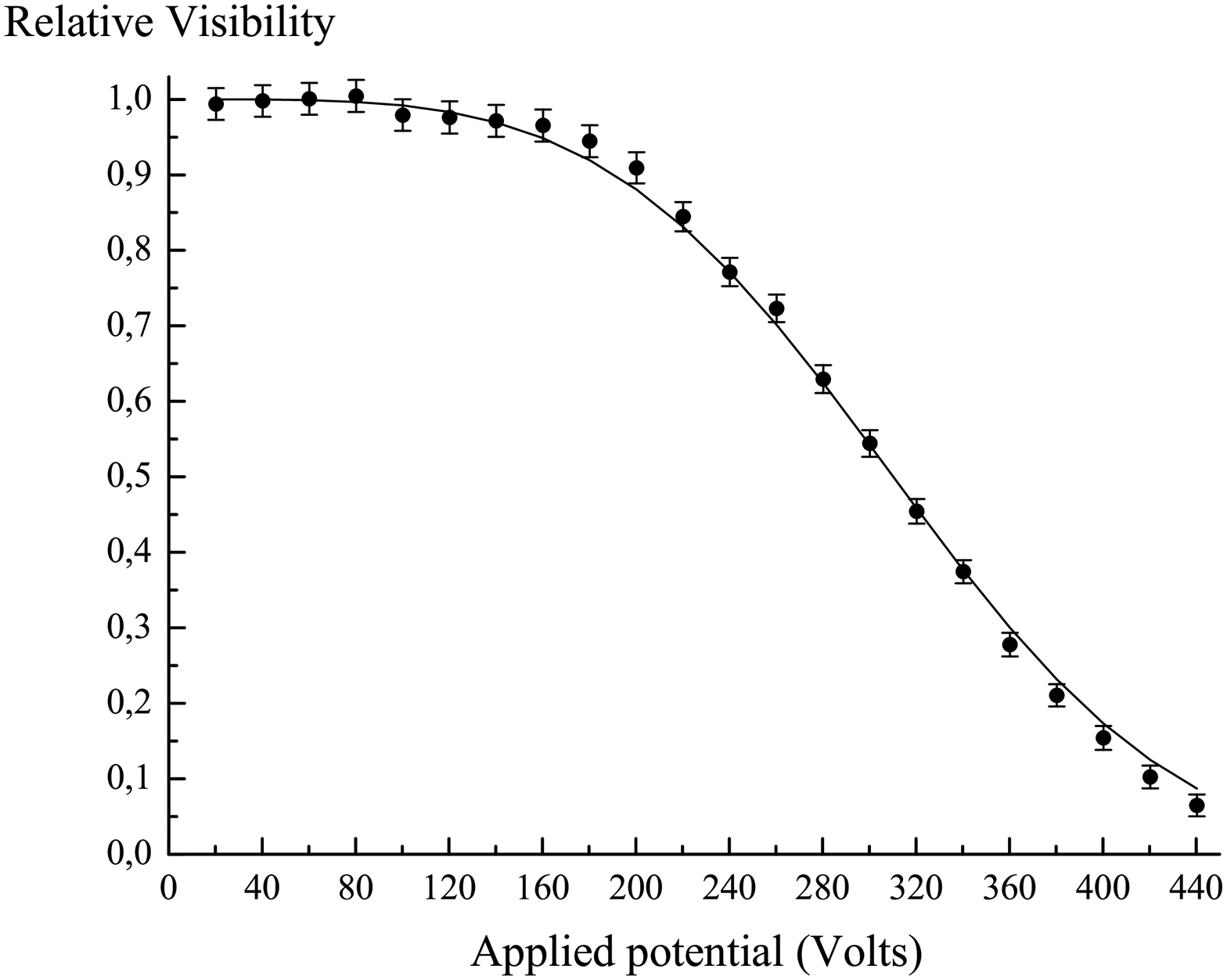}
\caption{\label{visibility} Relative fringe visibility
${\mathcal{V}}/{\mathcal{V}}_0$ (with ${\mathcal{V}}_0 = 62$\%) as
a function of the applied voltage $V_0$ and the best fit using
equations (\ref{n4},\ref{n5}) (full curve).}
\end{figure}

The phase shift $\phi(V_0)$ due to the polarizability effect is
taken equal to $\phi(V_0) = \left< \psi_i \right> - \left(\left<
\psi_{i-1} \right> + \left< \psi_{i+1} \right>\right)/2 $ where
the recording $i$ corresponds to the applied voltage $V_0$: the
average of the mean phase of the two $V_0=0$ recordings done just
before and after is our best estimator of the mean phase of the
interference signal in zero field. In figures \ref{phaseshifts}
and \ref{visibility}, we have plotted the phase shift $\phi(V_0)$
and the fringe visibility ${\mathcal{V}}$ as a function of the
applied voltage $V_0$.

To interpret these results, we must take into account the velocity
distribution of the lithium atoms, as the phase shift is
proportional to $v^{-1}$. We assume that the velocity distribution
is given by:

\begin{equation}
\label{n4} P(v)  = \frac{S_{\|}}{u \sqrt{\pi}} \exp\left[-\left(
(v-u)S_{\|}/u\right)^2\right]
\end{equation}
\noindent with the most probable velocity $u$ and $S_{\|}$ is the
parallel speed ratio. The traditional $v^3$ pre-factor
\cite{haberland85}, which has minor effects, is omitted when
$S_{\|}$ is large. The interference signals $I$ can be written:

\begin{equation}
\label{n5} I = I_0 \int dv P(v)  \left[1 + {\mathcal{V}}_0
\cos\left( \psi + \phi_m  \frac{u}{v} \right) \right]
\end{equation}

\noindent with $\phi_m = \phi(u)$. If we expand $u/v$ in powers of
$(v-u)/u$ up to second order, the integral can be calculated
exactly. This approximation is very good \cite{miffre05a} but not
accurate enough and we have calculated the integral (\ref{n5})
numerically. We have thus made a single fit for the phase and
visiblility results, with two adjustable parameters:
$\phi_m/V_0^2$ and $S_{\|}$. As shown in figures \ref{phaseshifts}
and \ref{visibility}, the agreement is very good, in particular
for $\left<\phi \right>$, and we deduce a very accurate value
$\phi_m/V_0^2$:

\begin{equation}
\label{n8} \frac{\phi_m}{V_0^2} = \frac{2 \pi \epsilon_0 \alpha
L_{eff}}{\hbar u \left<h \right>^2} = (1.3870  \pm 0.0010)\times
10^{-4} {\mbox{   rad/V}}^2\end{equation}

\noindent The relative uncertainty $0.07$\% proves the coherence
of our measurements. The parallel speed ratio $S_{\|} = 8.00 \pm
0.06$ is slightly larger than expected for our lithium beam,
because Bragg diffraction is velocity selective.

From measurements made on our capacitor, we get $2a = 50.00 \pm
0.10$ mm and $\left<h \right> = 2.056\pm 0.003$ mm. We have
measured the mean velocity $u$ using Doppler effect, by recording
atom deflection due to photon recoil with a laser beam almost
counterpropagating with the atoms. The uncertainty on the cosine
of the angle is negligible ($0.12$\%) and we get $ u = 1066.4 \pm
8.0$ m/s. We have also recorded the diffraction probability as a
function of the Bragg angle, by tilting the mirror forming a
standing wave. Using an independent calibration of the mirror
rotation as a function of the applied voltage on the piezo-drive,
we get a measurement of the Bragg angle $ \theta_B = h/(mu
\lambda_L) = 79.62 \pm 0.63 $ $\mu$rad corresponding to $u=1065.0
\pm 8.4$ m/s. These two values are perfectly coherent and we take
the mean velocity as their weighted average $u = 1065.7 \pm 5.8$
m/s. The theory of supersonic expansion can be used to check this
result: the velocity of a pure argon beam given by $u = \sqrt{5
k_BT_0/m}$ (where $T_0 = 1073 \pm 11$ K is the nozzle temperature
and $m$ the argon atomic mass) must be corrected: the dominant
correction is the velocity slip effect estimated to be $1$\%
\cite{skovorodko04a} and we get $u = 1068.4 \pm 5.5$ m/s in very
good agreement with our measurements.

Finally, we get the lithium electric polarizability of $^7$Li
$\alpha =(24.33 \pm 0.16)\times10^{-30} $ m$^3$, in excellent
agreement with the previous measurements, $\alpha = (22. \pm
2.)\times10^{-30}$ m$^3$ by Chamberlain and Zorn
\cite{chamberlain63} in 1963 and $\alpha = (24.3 \pm
0.5)\times10^{-30}$ m$^3$, by Bederson and co-workers
\cite{molof74} in 1974. Our result compares also very well with ab
initio calculations of $\alpha$: most calculations predict
$\alpha$ values in the range $(24.32-24.45)\times10^{-30}$ m$^3$
(see reference \cite{kobayashi97} and references therein).

With respect to the experiment done on sodium by D. Pritchard and
co-workers \cite{ekstrom95}, we have made several important
improvements:

Our capacitor design provides an analytical calculation of the
$E^2$ integral along the atomic path. This property is helpful in
minimizing the uncertainty on this quantity, through a better
understanding of the influence of small defects. With an improved
construction, we expect to reduce the uncertainty on this integral
below $0.1$ \%, the main limitation being due to the unknown
potential in the dielectric gaps.

Thanks to a large signal and an excellent fringe visibility, the
phase sensitivity of our interferometer is considerably larger
than previously achieved. The accuracy on phase measurement is
presently limited by the lack of reproducibility of the mean phase
of the recordings. We hope to improve this reproducibility by
stabilizing the temperature of the rail supporting the three
mirrors. The consistency and accuracy of our phase measurements is
proved by the quality of the fit of figure \ref{phaseshifts} and
by the $0.07$\% uncertainty obtained for the measurement of
$\phi_m/V_0^2$. We have deduced the value of the electric
polarizability $\alpha$ with a $0.66$\% relative uncertainty
dominated by the uncertainty on the mean atom velocity $u$.

Our interferometer is species selective thanks to laser
diffraction and this is also a very favorable circumstance. In his
thesis, T. D. Roberts reanalyzes the measurement of sodium atom
electric polarizability made by C. R. Ekstrom et al.
\cite{ekstrom95}: he estimates that a weak contribution of sodium
dimer to the interference signals might have introduced a non
negligible systematic error in the result.

T. D. Roberts et al. \cite{roberts04} have devised a very clever
technique to correct for the velocity dependence of the phase
shift $\Delta\phi$, so that they can observe fringes with a good
visibility up to very large $\phi$ values. The present result
proves that a very accurate measurement can be also made in the
presence of an important velocity dispersion without any
compensation of the associated phase dispersion, provided that the
velocity distribution is taken into account in the analysis.

Finally, we would like to emphasize two very striking properties
of atom interferometry. Our phase measurement consists in
measuring the increase $\Delta v$ of the atom velocity $v$ when
entering the field:

\begin{equation}
\label{n9} \frac{\Delta v}{v} = \frac{\lambda_{dB}}{L_{eff}}
\times\frac{ \phi }{2 \pi}
\end{equation}

\noindent $\Delta v/v$ is extremely small reaching only $\Delta
v/v \approx 4\times 10^{-9}$ for our largest field. Our ultimate
sensitivity, close to a $3$ mrad phase shift, corresponds to
$\Delta v/v \approx 6 \times 10^{-13}$!

\noindent In the capacitor, the atom wavefunction samples two
regions of space separated by $\sim 100$ $\mu$m  with a
macroscopic object lying in between and this situation extends
over $10^{-4}$ second, without any loss of coherence. This
consequence of quantum mechanics remains surprising!

We thank J. C. Lehmann and J. F. Arribart from Saint-Gobain, J. F.
Bobo and M. Nardone from LPMC and the staff of the AIME for their
help in the construction of the capacitor. We thank CNRS SPM and
R\'egion Midi Pyr\'en\'ees for financial support.


\end{document}